\documentclass[12pt,elegant]{article}
\usepackage{graphicx}
\usepackage{latexsym}
\usepackage{a4wide}
\usepackage{amsmath}
\usepackage{amsfonts,amssymb}

\newcommand{\lin}{l_{in}}
\newcommand{\lout}{l_{out}}
\newcommand{\genus}{{\mathfrak g}}
\newcommand{\dL}{\frac{\partial}{\partial L}}
\newcommand{\ddL}{\frac{\partial^2}{\partial L^2}}
\newcommand{\myref}[1]{(\ref{#1})}
\newcommand{\expec}[1]{\left<#1\right>}

\newcommand{\mi}{\!-\!}
\newcommand{\equ}{\!=\!}
\newcommand{\pl}{\!+\!}

\title{
{\normalsize \hfill PITHA-05/06}\\
\vspace{-1.5cm}
{\normalsize \hfill ITP-UU-05/26}\\
\vspace{-1.5cm}
{\normalsize \hfill SPIN-05/20}\\
${}$\\ 
${}$\\ 
${}$\\ 
\LARGE Taming the cosmological constant in 2D causal quantum gravity with
topology change\\
              }
\author{{\large         R. Loll$^{\dag}$, W. Westra$^{\dag}$ and S. Zohren$^{\dag,\ddag}$}\\[10pt]
        {\footnotesize \em \dag Institute for Theoretical Physics, Utrecht University}\\[-5pt]
        {\footnotesize \em Leuvenlaan 4, NL-3584 CE Utrecht, The Netherlands}\\
        {\footnotesize \em \ddag Institut f\"ur Theoretische Physik E, RWTH-Aachen}\\[-5pt]
        {\footnotesize \em D-52056 Aachen, Germany}\\[5pt]
        {\footnotesize \em E-Mail:}\\[-5pt]
        {\footnotesize  r.loll@phys.uu.nl, w.westra@phys.uu.nl and zohren@physik.rwth-aachen.de}
        }
\date{}         
        
\begin{document}

\maketitle \vspace*{1cm}

\begin{abstract} As shown in previous work, there is a well-defined nonperturbative
gravitational path integral including an explicit sum over topologies 
in the setting of Causal Dynamical Triangulations in two dimensions.
In this paper we derive a complete analytical solution of the quantum
continuum dynamics of this model, obtained uniquely by means of a
double-scaling limit. We show that the presence of infinitesimal wormholes 
leads to a decrease in the effective cosmological constant, reminiscent of
the suppression mechanism considered by Coleman and others in the 
four-dimensional Euclidean path integral.
Remarkably, in the continuum limit we obtain a finite spacetime density of
microscopic wormholes without assuming fundamental discreteness. 
This shows that one can in principle make
sense of a gravitational path integral which includes a sum over
topologies, provided suitable causality restrictions are imposed
on the path integral histories.
\end{abstract}
\pagebreak

%  ----------------- Introduction -----------------

\section{Introduction}

Despite recent progress \cite{spec,univ}, little is known about the ultimate
configuration space of quantum gravity on which its nonperturbative
dynamics takes place. This makes it difficult to decide 
which (auxiliary) configuration space to choose as starting
point for a quantization. In the context of a path integral
quantization of gravity, the relevant question is which class of
geometries one should be integrating over in the first place. 
Setting aside the formidable difficulties in ``doing the integral", there is
a subtle balance between including too many geometries -- such
that the integral will simply fail to exist (nonperturbatively) in any
meaningful way, even after renormalization -- and including too few
geometries, with the danger of not capturing a physically relevant part
of the configuration space. 

A time-honoured part of this discussion is the question of whether a 
sum over different spacetime topologies should be included in the 
gravitational path integral. The absence to date of a viable theory of
quantum gravity in four dimensions has not hindered speculation on
the potential physical significance of processes involving topology
change (for reviews, see \cite{horo,dowker}). Because such processes 
necessarily violate causality, they are usually considered
in a Euclidean setting where the issue does not arise. Even if one
believes that Euclidean quantum gravity {\it without} a sum over topologies
exists nonperturbatively as a fundamental theory of nature -- something
for which there is currently little evidence --, insisting on including such a sum  
makes it all the more difficult to perform the path integral and extract
physical information from it. This happens because the 
number of ways in
which one can cut and reglue a manifold to obtain manifolds with
a different topology is very large and leads to uncontrollable divergences
in the path integral.\footnote{Even in the simplest case of two-dimensional
geometries, the number of possible configurations 
grows faster than exponentially with the volume of the geometry.
A well-known manifestation of this problem is the non-Borel summability of the genus
expansion in string theory. This does not necessarily mean that there is
no underlying well-defined theory, but even in the much-studied case of
two-dimensional Euclidean quantum gravity no physically satisfactory, 
unambiguous solution has been found \cite{DiFrancesco:1993nw}.}
  
Formal path integrals involving nontrivial topologies are either semiclassical
and assume that a (usually small and hand-picked) class of such configurations
dominates the path integral, without being able to check that they are indeed
saddle points of a full, nonperturbative formulation (see \cite{sh} for a
recent example), or otherwise 
postulate a separation of scales between the
fundamental quantum excitations of the geometry and a (larger) length
scale characteristic for the topology changes, for example, the size
of wormholes \cite{Coleman:1988tj}. We are not aware of any evidence
from nonperturbative approaches that would support either of these
assumptions. On the contrary, as already mentioned, path integrals
including a sum over topologies tend not to exist at all. Furthermore, 
as is clear from studies of both Euclidean \cite{DiFrancesco:1993nw}
and Lorentzian \cite{Ambjorn:1999fp} sums over
geometries in two dimensions, once topology changes are allowed 
dynamically, they occur everywhere and at all scales.\footnote{In the
Lorentzian case cited, the standard sum over causal geometries
of a {\it fixed} spacetime topology \cite{2dlor} is extended by allowing ``baby universes"
which branch off the main universe, causing changes in the topology
of spatial slices as a function of time, but not changing the 
spacetime topology.}

A new idea to tame the divergences associated with topology
changes in the path integral was advanced in \cite{ourtopology1}
and implemented in a model of two-dimensional nonperturbative 
Lorentzian quantum
gravity. The idea is to include a sum over topologies, or over some
subclass of topologies, in the state sum, but to restrict this class further
by certain {\it geometric} (as opposed to topological) constraints.
These constraints involve the causal (and therefore Lorentzian)
structure of the spacetimes and thus would have no analogue in a purely
Euclidean formulation. In the concrete two-dimensional model
considered in \cite{ourtopology1}, the path integral is taken over a
geometrically distinguished class of spacetimes with arbitrary numbers
of ``wormholes", which violate causality only relatively mildly (see
also \cite{ourtopology2}). As a consequence, the nonperturbative
path integral turns out to be well defined. 
This is an extension of the central idea of the approach of causal
dynamical triangulations, namely, to use physically motivated 
causality restrictions to make the gravitational path integral better behaved
(see \cite{myrev} for a review). 

In this paper, we will present a complete analytical solution of the
statistical model of two-dimensional Lorentzian random geometries
introduced in \cite{ourtopology1}, whose starting point is a regularized
sum over causal triangulated geometries {\it including} a sum over
topologies. For a given genus (i.e. number of (worm)holes in the spacetime)
not all possible triangulated geometries are included in the sum,
but only those which satisfy certain causality constraints. As shown
in \cite{ourtopology1}, this makes the statistical model well defined,
and an unambiguous continuum limit is obtained by taking a
suitable double-scaling limit of the two coupling constants of the model,
the gravitational or Newton's constant and the cosmological
constant. The double-scaling limit presented here differs from the
one found in \cite{ourtopology1,ourtopology2}, where only the partition
function for a single spacetime strip was evaluated. We will show that
when one includes the boundary lengths of the strip explicitly --
as is necessary to obtain the full spacetime dynamics -- the natural
renormalization of Newton's constant involves the boundary
``cosmological" coupling constants conjugate to the boundary lengths. 
Although the holes we include exist only for an infinitesimal time, and we
do not keep track of them explicitly in the states of the Hilbert
space, their integrated effect is manifest in the continuum Hamiltonian 
of the resulting gravity theory. As we will see, their presence leads
to an effective lowering of the cosmological constant and therefore
represents a concrete and nonperturbative implementation of an idea 
much discussed in the late eighties in the context of the ill-defined 
continuum path integral formulation of Euclidean quantum gravity  
(see, for example, \cite{Coleman:1988tj,Klebanov:1988eh}).

The remainder of the paper is structured as follows. In the next section,
we briefly describe how a nonperturbative theory of two-dimensional
Lorentzian quantum gravity can be obtained by the method of causal
dynamical triangulation (CDT), and how a sum over topologies can be
included. For a more detailed account of the construction of
topology-changing spacetimes and the
geometric reasoning behind the causality constraints
we refer the reader to \cite{ourtopology1,ourtopology2}.
The main result of Sec.\ \ref{discretesection} is the computation
of the Laplace transform of the one-step propagator of the discrete 
model for arbitrary boundary geometries.
In Sec.\ \ref{continuumsection} we make a scaling ansatz for the 
coupling constants and show that just one of the choices for
the scaling of Newton's constant leads to a new and physically
sensible continuum theory. We calculate the corresponding quantum
Hamiltonian and its spectrum, as well as the full propagator of
the theory. Using these results, we compute
several observables of the continuum theory in Sec.\ \ref{sec_observ},
most importantly, the expectation
value of the number of holes and its spacetime density. 
In Sec.\ \ref{conclusions}, we summarize
our results and draw a number of conclusions. In Appendix A, we
discuss the properties of alternative scalings for Newton's constant
which were discarded in the main text. This also establishes a
connection with a previous attempt \cite{DiFrancesco:2000nn} 
to generalize the original Lorentzian model without topology
changes. In Appendix B, we calculate the spacetime density
of holes from a single infinitesimal spacetime strip.

%  ----------------- Lorentzian Sum over Topologies -----------------
\section{Lorentzian sum over topologies}\label{lorentzianchapter}

Our aim is to calculate the (1+1)-dimensional
gravitational path integral 
\begin{equation}\label{contPI}
    Z(G_N,\Lambda) = \sum_{topol.} \int D[g_{\mu\nu}] e^{i
    S(g_{\mu\nu})}
\end{equation}
nonperturbatively by using the method of Causal Dynamical
Triangulations (CDT).\footnote{For an introduction to CDT the 
reader is referred to \cite{2dlor,2dlor2, ajl4d}.}.
The sum in (\ref{contPI}) denotes the inclusion in the path integral
of a specific, causally preferred class of
fluctuations of the manifold topology. The action $S(g_{\mu\nu})$
consists of the usual Einstein-Hilbert curvature term and   
a cosmological constant term. Since we work in two dimensions, the 
integrated curvature term is proportional to the Euler characteristic
$\chi = 2\mi 2\genus\mi b$ of the spacetime manifold, where $\genus$ 
denotes the genus (i.e. the
number of handles or holes) and $b$ the number of boundary components. 
Explicitly, the action reads
\begin{equation}\label{action}
    S = 2 \pi \chi K - \Lambda \int d^2x\sqrt{|\det g_{\mu\nu}|} 
\end{equation}
where $K=1/G_N$ is the inverse Newton's constant and $\Lambda$ the
cosmological constant (with dimension of inverse length squared).

\begin{figure}
\begin{center}
\includegraphics[width=5in]{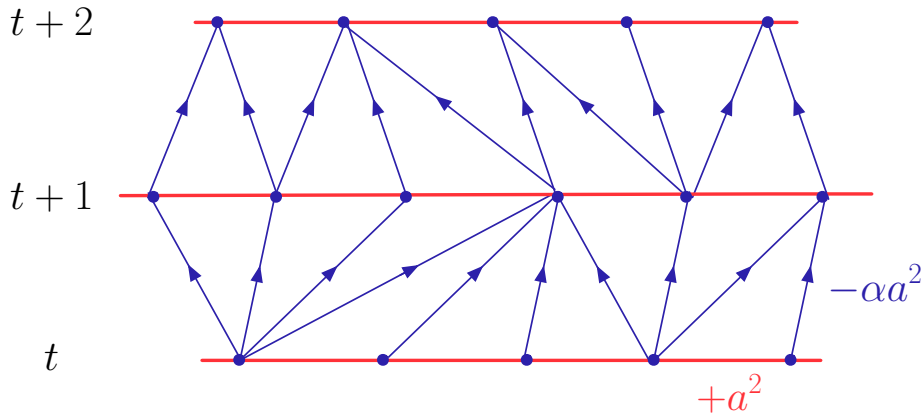}
\caption{A section of a sequence $[t,t+2]$ of two spacetime strips of a
triangulated two-dimensional spacetime contributing to the regularized path 
integral without
topology changes.}\label{figure01}
\end{center}
\end{figure}

Just like in the original CDT model \cite{2dlor} (which from now on we will
also refer to as the ``pure" model, i.e. without topology changes), we will 
first regularize the path integral \myref{contPI} by a sum 
over piecewise flat two-dimensional spacetimes, whose flat building
blocks are identical Minkowskian triangles, all with 
one space-like edge of squared length $+a^2$ and two time-like edges of 
squared length $-\alpha a^2$, where $\alpha$ is a real positive constant. 
The CDT path integral takes the form of a sum over triangulations, 
with each triangulation consisting of a sequence of spacetime strips of height 
$\Delta t =1$ in the time direction. A single such strip is a set of  $l_{in}$ 
triangles pointing up and $l_{out}$ triangles pointing down (Figure \ref{figure01}). 
Because the geometry has a sliced structure, one can easily Wick-rotate it
to a triangulated manifold of Euclidean signature by analytically continuing
the parameter $\alpha$ to a real negative value \cite{ajl4d}. For simplicity, we will set
$\alpha\equ -1$ in evaluating the regularized, real and Wick-rotated version of
the path integral (\ref{contPI}).

\begin{figure}
\begin{center}
\includegraphics[width=3.5in]{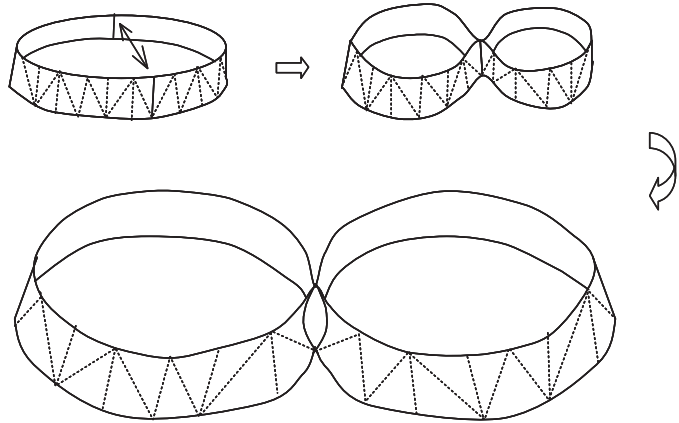}
\caption{Construction of a wormhole by identifying two time-like
edges of a spacetime strip and cutting open the geometry along the
edge.}\label{figure02}
\end{center}
\end{figure}

In the pure CDT model the one-dimensional spatial slices of constant 
proper time $t$ are usually chosen as circles, resulting 
in cylindrical spacetime geometries. For our present purposes, we
will enlarge this class of geometries by allowing the genus to be variable.
We define the sum over topologies by
performing surgery moves directly on the triangulations to obtain
regularized versions of higher-genus manifolds \cite{ourtopology1,ourtopology2}. 
They are generated by adding tiny wormholes that connect two regions of 
the same spacetime strip. Starting from a regular strip of
topology $[0,1]\times S^1$ and height $\Delta t =1$, one can construct a hole by 
identifying two of the strip's time-like edges and subsequently cutting open the 
geometry along this edge (Fig.\ \ref{figure02}). By applying this procedure 
repeatedly (obeying certain causality constraints \cite{ourtopology1,ourtopology2}),
more and more wormholes can be created. 
Once the regularized path integral has been performed, including a sum over
geometries with wormholes, one takes a continuum limit by letting $a\rightarrow 0$ 
and renormalizing the coupling constants appropriately, as will be described
in the following sections. 

Note that our wormholes are minimally causality- and locality-violating
in that they are located within a single proper-time step (the smallest time
unit available in the discretized theory) and the associated baby universes
which are born at time $t$ are reglued at time $t+1$ ``without twist" \cite{ourtopology1,ourtopology2}.
In a macroscopic interpretation one could describe them as wormholes
which are instantaneous in the proper-time frame of an ensemble of
freely falling observers. Note that this is invariantly defined (on Minkowski
space, say) once an initial surface has been chosen. Such a restriction is
necessary if one wants to arrive at a well-defined unitary evolution
via a transfer matrix formalism, as we are doing. To include wormholes
whose ends lie on different proper-time slices, one would have to
invoke a third-quantized formulation, which would very likely result in
{\it macroscopic} violations of causality, locality and therefore unitarity,
something we are trying to avoid in the present model.

One could wonder whether the effect in the continuum theory of our choice
of wormholes is to single out a preferred frame or coordinate system. Our
final result will show that this is not the case, at least not over and above
that of the pure gravitational model without topology changes. The effect
of the inclusion of wormholes turns out to be a rather mild ``dressing" of the
original theory without holes. We believe that the essence of our model lies
not so much in how the wormholes are connected, because they do not
themselves acquire a nontrivial dynamics in the continuum limit.
Rather, it is important that their number is sufficiently large to have an effect
on the underlying geometry, but on the other hand sufficiently
controlled so as to render the model computable.

A similar type of wormhole has played a prominent role in past attempts to 
devise a mechanism to explain the smallness of the cosmological constant 
in the Euclidean path integral formulation of {\it four}-dimensional quantum gravity 
in the continuum \cite{Coleman:1988tj,Klebanov:1988eh}. 
The wormholes considered there resemble those of our toy model 
in that both are non-local identifications of the spacetime geometry of 
infinitesimal size. The counting of our wormholes is of course different since 
we are working in a genuinely Lorentzian setup where certain causality 
conditions have to be fulfilled. This enables us to do the sum over
topologies completely explicitly. Whether a similar construction is
possible also in higher dimensions is an interesting, but at this stage 
open question.

%  ----------------- Discrete Solution: The Transfer Matrix -----------------
\section{Discrete solution: the one-step propagator}\label{discretesection}

For the (1+1)-dimensional Lorentzian gravity model including a sum over
topologies, the partition function of a single spacetime strip of infinitesimal
duration with summed-over boundaries was evaluated in
\cite{ourtopology1} and \cite{ourtopology2}.
In the present paper, we will extend this treatment by calculating the 
full one-step propagator, or, equivalently, the generating function for the 
partition function of a single strip with given, fixed boundary lengths. 
This opens the way for investigating the full dynamics of the model.

The discrete set-up described above leads to the Wick-rotated one-step 
propagator 
\begin{equation}\label{Transferlaplacedef}
G_{\lambda,\kappa}(l_{in},l_{out},t=1)=
e^{-\lambda (\lin+\lout)}\sum_{T|l_{in},l_{out}}e^{-2\kappa\genus},
\end{equation}
where $\kappa$ is the bare inverse Newton's constant and $\lambda$ the bare (dimensionless)
cosmological constant, and we have omitted an overall constant coming from the 
Gauss-Bonnet integration. The sum in \myref{Transferlaplacedef} is to be taken over all 
triangulations with $l_{in}$ space-like links in the initial and $l_{out}$ space-like links in the 
final boundary.
Note that the number of holes does not appear as one of the arguments of the one-step 
propagator since we only consider holes that exist within one strip. 
Consequently, the number of holes does not appear explicitly as label for the quantum 
states, and the Hilbert space coincides with that of the pure CDT model. 
Nevertheless, the integrated effect of the topologically non-trivial
configurations changes the dynamics and the quantum Hamiltonian, as we shall see.

The one-step propagator \myref{Transferlaplacedef} defines a transfer matrix $\hat T$
by
\begin{equation}
\label{transfer}
G_{\lambda,\kappa}(l_{in},l_{out},1)=\langle l_{out}|\hat{T}| l_{in}\rangle,
\end{equation}
from which we obtain the propagator for $t$ time steps as usual by
iteration,
\begin{equation}
G_{\lambda,\kappa}(l_{in},l_{out},t)=\langle l_{out} | \hat{T}^t | l_{in} \rangle.
\label{fullprop}
\end{equation}
For simplicity we perform the sum in \myref{Transferlaplacedef} over triangulated
strips with periodically identified boundaries in the spatial direction and one 
marked time-like edge. By virtue of the latter,
$G_{\lambda,\kappa}(l_{in},l_{out},t)$ satisfies the desired composition
property of a propagator,
\begin{eqnarray}
\label{composition}
G_{\lambda,\kappa}(l_{in},l_{out},t_1+t_2)&=& \sum_{l}G_{\lambda,\kappa}(l_{in},l,t_1)
G_{\lambda,\kappa}(l,l_{out},t_2),\\
G_{\lambda,\kappa}(l_{in},l_{out},t+1)&=& \sum_{l}G_{\lambda,\kappa}(l_{in},l,1)
G_{\lambda,\kappa}(l,l_{out},t),
\end{eqnarray}
where the sums on the right-hand sides are performed over an intermediate
constant-time slice of arbitrary discrete length $l$.

Performing the fixed-genus part of the sum over triangulations in
\myref{Transferlaplacedef} yields
\begin{equation}
\label{sumfixedg} 
G_{\lambda,\kappa}(l_{in},l_{out},1)=e^{-\lambda 
N}\sum_{\genus=0}^{[N/2]}
\binom{N}{\lin}\binom{N}{2\genus}\frac{(2\genus)!}{\genus !
(\genus+1)!}e^{-2\kappa\genus},
\end{equation}
with $N \equ \lin \pl \lout$. To simplify calculations we will use
the generating function for\-malism with
\begin{equation}
\label{gen} 
G(x,y,g,h,1)=\sum_{\lin,\lout=0}^\infty
G_{\lambda,\kappa}(l_{in},l_{out},1) x^{\lin} y^{\lout},
\end{equation}
where we have defined $g\equ e^{-\lambda }$ and
$h\equ e^{-\kappa}$. The quantities $x$ and $y$ can be seen
as purely technical devices, or alternatively as exponentiated
bare boundary cosmological constants
\begin{equation}
%\label{ } 
x=e^{- \lambda_{in}},\quad y=e^{- \lambda_{out}}.
\end{equation}
Upon evaluating the
sum over $\lin$ and $\lout$ one obtains the generating
function of the one-step propagator
\begin{equation}
\label{Glaplace}
 G(x,y,g,h,1)=    \frac{1}{1 - g\,\left( x + y \right) }
 \frac{2}{1 + {\sqrt{1 - 4 u^2}}},
\end{equation}
with
\begin{equation} \label{z}
  u = \frac{h}{ \frac{1}{g\,\left( x + y \right) }-1}.
\end{equation}
Note that in order to arrive at the final result (\ref{Glaplace}),
we have performed an explicit sum over all topologies! 
The fact that this infinite sum converges for appropriate values of
the bare couplings has to do with the causality constraints 
imposed on the model, which were geometrically motivated
in \cite{ourtopology1}, and which effectively reduce the number of
geometries in the genus expansion.

In \myref{Glaplace} one recognizes the generating function 
${\rm Cat}(u^2)$ for the Catalan numbers,
\begin{equation}
\label{catalan}
{\rm Cat}(u^2)=\frac{2}{1 + {\sqrt{1 - 4 u^2}}}.
\end{equation}
For $h\equ 0$ one has ${\rm Cat}(u^2)\equ 1$ and expression 
(\ref{Glaplace}) reduces to the
one-step propagator without topology changes,
\begin{equation}
\label{transfernog}
G(x,y,g,h=0,1)=\frac{1}{1 - g\,\left( x + y \right) }.
\end{equation}
Furthermore, one recovers the one-step partition function with
summed-over boundaries of  \cite{ourtopology1,ourtopology2} by 
setting $x\equ y\equ 1$,
\begin{equation}
\label{Znog}
Z(g,h,1) = \frac{1}{1 - 2 g }
\frac{2}{1 + {\sqrt{1 - 4 (\frac{2g h}{1-2g})^2}}}.
\end{equation}

%  ----------------- The Continuum Limit -----------------
\section{Taking the continuum limit}\label{continuumsection}

Taking the continuum limit in the case without topology changes is
fairly straightforward \cite{2dlor}. The joint region of
convergence of \myref{transfernog} is given by
\begin{equation}
\label{convergence }
|x|<1,\quad |y|<1,\quad |g|<\frac{1}{2}.
\end{equation}
One then tunes the couplings to their critical values according to
the scaling relations
\begin{eqnarray}
g = \frac{1}{2}(1-&\!\!\!\!\!\!\!\!\! &a^2\,\Lambda) + \mathcal{O}(a^{3})\label{scalinga}, \\
x  =  1-a\,X + \mathcal{O}(a^{2}),& \!\!\!\!\!\!\!\!\!\!\! &\quad %\nonumber,\\
y  =  1-a\,Y + \mathcal{O}(a^{2})\label{scalingb}.
\end{eqnarray}
Up to additive renormalizations, $x$, $y$ and $\lambda$ scale canonically, 
with corresponding renormalized couplings $X$, $Y$ and $\Lambda$. 
In the case with
topology change we have to introduce an additional scaling
relation for $h$. Since Newton's constant is dimensionless in two
dimensions,
there is no preferred canonical scaling for $h$. 
%For dimensional reasons 
We make the multiplicative ansatz\footnote{Here the factor
$\frac{1}{\sqrt{2}}$ is chosen to give a proper parametrization of
the number of holes in terms of Newton's constant (see Section
\ref{sec_observ}).}
\begin{equation}
\label{scalingh} h =\frac{1}{\sqrt{2}} h_{ren} (a d)^\beta,
\end{equation}
where $h_{ren}$ depends on the renormalized Newton's constant
$G_N$ according to 
\begin{equation}
\label{hrendef}
h_{ren}=e^{- 2\pi / G_N}.
\end{equation}
In order to compensate the powers of the cut-off $a$ in (\ref{scalingh}),
$d$ must have dimensions of inverse length.
The most natural ansatz in terms of the dimensionful quantities 
available is
\begin{equation} 
\label{d}
d = (\sqrt{\Lambda}^{\alpha} (X+Y)^{1-\alpha}).
\end{equation}
The constants $\beta$ and $\alpha$ in relations (\ref{scalingh})
and (\ref{d}) must be chosen such as to
obtain a physically sensible continuum theory. By
this we mean that the one-step propagator should
yield the Dirac delta-function to lowest order in $a$, and that the
Hamiltonian should be bounded below and not
depend on higher-order terms in \myref{scalinga}, \myref{scalingb},
in a way that would introduce a dependence on new couplings without
an obvious physical interpretation. 

To calculate the Hamiltonian operator $\hat H$ we use the analogue of the 
composition law \myref{composition} for the Laplace
transform of the one-step propagator \cite{2dlor},
\begin{equation}
\label{laplacecomposition}
G(x,y,t+1)= \oint \frac{dz}{2 \pi i z} G(x,z^{-1};1)G(z,y,t).
\end{equation}
In a similar manner we can write the time evolution of the wave function as
\begin{equation}
\label{timeevolution1}
\psi(x,t+1)= \oint \frac{dz}{2 \pi i z} G(x,z^{-1};1)\psi(z,t).
\end{equation}
When inserting the scaling relations \myref{scalinga}, \myref{scalingb}
and $t\equ \frac{T}{a}$ into this equation it is convenient to treat separately the
first factor in the one-step propagator \myref{Glaplace}, which is nothing but the
one-step propagator without topology changes \myref{transfernog}, and
the second factor, the Catalan generating function \myref{catalan}. 
Expanding both sides of \myref{timeevolution1} to order $a$ gives
\begin{equation}
\label{scaledtransfernog}
\left(1-a \hat{H}+\mathcal{O}(a^2)\right)\psi(X) = \int^{i\infty}_{-i\infty} 
\frac{dZ}{2\pi i} \left\{\left(\frac{1}{Z-X} + a\,\frac{2\Lambda - X Z}{(Z-X)^2}\right)
{\rm Cat}(u^2) \right\} \psi(Z),
\end{equation}
where we have used
\begin{equation}
\psi(X,T+a)=e^{-a\,\hat{H}}\psi(X,T),
\end{equation}
with $\psi(X)\equiv\psi(x\equ 1- a X)$.
Note that the first term on the right-hand side of \myref{scaledtransfernog}, $\frac{1}{Z-X}$, 
is the Laplace-transformed delta-function. The interesting new behaviour of the Hamiltonian 
is contained in the expansion of the Catalan generating function.
Combining \myref{catalan} and \myref{z}, and inserting the scalings
\myref{scalinga}, \myref{scalingb}, yields
\begin{equation} 
\label{scalingcatalan}
{\rm Cat}(u^2) = 1 + { \frac{2\,{d }^{2\beta }\,h_{ren}^2}{\,{\left( Z-X \right) }^2}
a^{2\beta-2} + {\rm h.o.} },
\end{equation}
where h.o. refers to terms of higher order in $a$.
In order to preserve the delta-function and have a
non-vanishing contribution to the Hamiltonian one is thus naturally led to
$\beta \equ 3/2$. For suitable choices of $\alpha$ it is also possible to obtain 
the delta-function by setting $\beta\equ 1$, but
the resulting Hamiltonians turn out to be unphysical or at least do not
have an interpretation as gravitational models with wormholes, 
as we will 
discuss in Appendix A.\footnote{One might also consider scalings of the form 
$ h \rightarrow  c_1 h_{ren} (a d)+c_2 h_{ren} (a d)^{3/2}$, but
they can be discarded by arguments similar to those of Appendix A.
}

For $\beta = 3/2$ the right-hand side of \myref{scaledtransfernog} becomes
\begin{equation} 
\label{scaledtransfernognew}
 \int^{i\infty}_{-i\infty} \frac{dZ}{2\pi i} \left\{\frac{1}{Z-X} + a\,\left( \frac{2\Lambda - X Z}
{(X-Z)^2}-\frac{2\,\sqrt{\Lambda}^{3 \alpha} \,h_{ren}^2}
{\,{\left( X-Z \right) }^{3 \alpha}} \right)\right\} \psi(Z).
\end{equation}
We observe that for $\alpha\leqslant 0$ the last term in \myref{scaledtransfernognew} 
does not contribute to the Hamiltonian.
Performing the integration for $\alpha>0$ and discarding the possibility of fractional poles 
the Hamiltonian reads
\begin{equation}
\label{HamiltonianX}
\hat{H}(X,\frac{\partial}{\partial X})=X^2 \frac{\partial}{\partial X} + 
X -2 \Lambda \frac{\partial}{\partial X}+
2\Lambda^{\frac{3 \alpha}{2}}h_{ren}^2
\frac{(-1)^{3\alpha}}{ \Gamma(3\alpha)} \frac{\partial^{3\alpha-1}}{\partial X^{3\alpha-1}},\quad \alpha=\frac{1}{3},\frac{2}{3},1,...
\end{equation}
For all $\alpha$'s, these Hamiltonians do not depend on higher-order terms in the
scaling of the coupling constants. One
can check this by explicitly introducing a term quadratic in $a$ (which can
potentially contribute to $\hat H$) in the
scaling relations \myref{scalingb}, namely,
\begin{eqnarray}
x &=& 1-a\,X +\frac{1}{2}\gamma\,a^2\,X^2+\mathcal{O}(a^3),\nonumber\\
y &= & 1-a\,Y
+\frac{1}{2}\gamma\,a^2\,Y^2+\mathcal{O}(a^3),\label{scalingbnew}
\end{eqnarray}
and noticing that \myref{HamiltonianX} does not depend on $\gamma$.
After making an inverse Laplace transformation $\psi(L)=\int_0^\infty d X e^{X\,L}\psi(X)$ 
to obtain a wave function in the ``position" representation (where it depends on the
spatial length $L$ of the universe),
and introducing $m=3\alpha-1$ the Hamiltonian reads
\begin{equation}
\label{Hamiltonian}
\hat{H}(L,\dL)=-L\ddL-\dL+2\,\Lambda\,L- \frac{2\,\Lambda^\frac{m+1}{2}h_{ren}^2 }
{ \Gamma(m+1)} L^m,\quad m=0,1,2,...\, .
\end{equation}
Since $\hat H$ is unbounded below for $m \geqslant 2$, we are left
with $m\equ 0$ and $m\equ 1$ as possible choices for the scaling. 
However, setting $m\equ 0$ merely has the effect of adding a constant term 
to the Hamiltonian, leading to a trivial phase factor for the wave function.
We conclude that the only new and potentially interesting model corresponds 
to the scaling with $m\equ 1$ and
\begin{equation}
\label{hm1}
h^2=\frac{1}{2} h_{ren}^2\,\Lambda\,(X+Y)\,a^3,
\end{equation}
with the Hamiltonian given by
\begin{equation}
\label{Hamiltonian1}
\hat{H}(L,\dL)=-L\ddL-\dL+  \left(1- h_{ren}^2 \right)\,2\,\Lambda\,L.
\end{equation}
Note that for all values $G_N\geq 0$ of the renormalized
Newton's constant \myref{scalingh} the Hamiltonian is bounded from
below and therefore well defined. It is self-adjoint
with respect to the natural measure $d\mu(L)\equ dL$ and has a discrete
spectrum, with eigenfunctions
\begin{equation}
\label{eigenfunc}
\psi_n(L)=\mathcal{A}_n e^{ -\sqrt{2\Lambda(1- h_{ren}^2)}L}
L_n(\, 2\sqrt{2\Lambda(1-h_{ren}^2)}\, ),\qquad n=0,1,2,...\, ,
\end{equation}
where $L_n$ denotes the $n$'th Laguerre polynomial. Choosing the normalization 
constants as
\begin{equation}
%\label{ }
\mathcal{A}_n=\sqrt[4]{8\Lambda(1- h_{ren}^2)},
\end{equation}
the eigenvectors $\{\psi_n(L)$, $n\equ 0,1,2,...\}$ 
form an orthonormal basis, and
the corresponding eigenvalues are given by
\begin{equation}
%\label{ }
E_n=\sqrt{2\Lambda(1- h_{ren}^2)}\left(2n+1\right),\quad n=0,1,2,...\, .
\end{equation}
Having obtained the eigenvalues one can easily calculate the
Euclidean partition function for finite time $T$ (with time periodically 
identified) 
\begin{equation}
\label{continuumpartfunc} Z_T(G_N,\Lambda)=\sum_{n=0}^\infty e^{-T\,E_n} =
\frac{e^{-\sqrt{2\Lambda(1- h_{ren}^2)}
T}}{1-e^{-2\,\sqrt{2\Lambda(1- h_{ren}^2)}T}},\quad h_{ren}=e^{- 2\pi / G_N}.
\end{equation}
For completeness we also compute the finite-time propagator 
\begin{equation}
    G_{\Lambda,G_N}(L_1,L_2,T) \equiv \langle{L_2} | e^{-T\hat{H}}| L_1\rangle =
%\\ &=& 
\sum_{n=0}^{\infty}e^{-TE_n}\psi_n^*(L_2)\psi_n(L_1)\label{FTPropdef}.
\end{equation}
Inserting \myref{eigenfunc} into \myref{FTPropdef} and using known relations 
for summing over Laguerre polynomials \cite{integrals} yields
\begin{eqnarray}\label{calogero20}
    G_{\Lambda,G_N}(L_1,L_2,T)=\omega\,
    \frac{e^{-\omega(L_1+L_2)\coth(\omega T)}}{\sinh(\omega T)}
     I_0 \left(\frac{2 \omega \, \sqrt{L_1 L_2} }{\sinh(\omega T)} \right),
\end{eqnarray}
where we have used the shorthand notation $\omega=\sqrt{2\Lambda(1-
h_{ren}^2)}$. As expected, for $h_{ren}\rightarrow 0$ the results
reduce to those of the pure two-dimensional CDT model.

\section{Observables} \label{sec_observ}

Due to the low dimensionality of our quantum-gravitational model,
it has only a few observables which characterize its physical properties.
Given the eigenfunctions \myref{eigenfunc} of the Hamiltonian \myref{Hamiltonian1} 
one can readily calculate the average spatial extension $\langle L\rangle$ of the
universe and all higher moments
\begin{equation}
\label{moments}
\expec{L^m}_n = \int_0^\infty dL\,L^m |\psi_n(L)|^2.
\end{equation}
Using integral relations for the Laguerre polynomials \cite{integrals} 
one obtains\footnote{Note that the poles of $\Gamma(-m)$ cancel with 
those of the hypergeometric function.}
\begin{eqnarray}
\expec{L^m}_n
= \left(\frac{1}{8\Lambda (1- h_{ren}^2)}\right)^\frac{m}{2}\frac{\Gamma(n-m)
\Gamma(m+1)}{\Gamma(n+1)\Gamma(-m)}\times\nonumber\\\nonumber\\
 \times\,\, {}_3F_2 (-n,1+m,1+m;1,1+m-n;1),
\end{eqnarray}
where ${}_3F_2 (a_1,a_2,a_3;b_1,b_2;z)$ is the generalized hypergeometric 
function defined by
\begin{equation}
\label{hypeerdefh}
{}_3F_2 (a_1,a_2,a_3;b_1,b_2;z)=\sum_{k=0}^\infty
\frac{(a_1)_k (a_2)_k (a_3)_k\,z^k}{(b_1)_k (b_2)_k k!}.
\end{equation}
Observe that the moments scale as $\expec{L^m}_n\sim \Lambda^{-\frac{m}{2}}$ 
which indicates that the effective Hausdorff dimension is given by $d_H=2$,
just like in the pure CDT model \cite{2dlor2}.
\begin{figure}
\begin{center}
\includegraphics[width=4in]{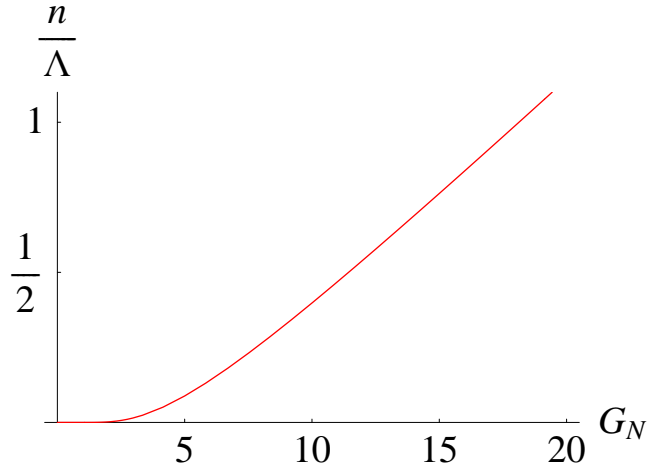}
\caption{The density of holes $n$ in units of $\Lambda$ as a
function of Newton's constant $G_N$.}\label{figure1}
\end{center}
\end{figure}

In addition to these well-known geometric observables, 
the system possesses a new type of ``topological" observable
which involves the number of holes $N_\genus$, as already anticipated in
\cite{ourtopology1,ourtopology2}. As spelled out there, the presence of 
holes in the quantum geometry and
their density can be determined from light scattering.
An interesting quantity to
calculate is the average number of holes in a piece of spacetime
of duration $T$, with initial and final spatial boundaries identified. 
Because of the simple dependence of the action on the genus
this is easily computed by taking the derivative of the
partition function $Z_T$ with respect to the corresponding
coupling, namely,
\begin{equation}
\label{Nholedefpartfunc}
    \expec{N_\genus} = \frac{1}{Z_T}\frac{h_{ren}}{2}\frac{\partial\, Z_T}{\partial h_{ren}}.
\end{equation}
Upon inserting \myref{continuumpartfunc} this yields
\begin{equation}\label{Nholeresultpartfunc}
    \expec{N_\genus} =  T \, h_{ren}^2 \Lambda \frac{\coth \left(\sqrt{2\Lambda
    (1-h_{ren}^2)}\,T\right)}{ \sqrt{2\Lambda (1-h_{ren}^2)}}.
\end{equation}
In an analogous manner we can also calculate the average spacetime volume
\begin{equation}\label{averagevoldef}
    \expec{V} = -\frac{1}{Z_T}\frac{\partial\, Z_T}{\partial \Lambda},
\end{equation}
leading to
\begin{equation}\label{averagevolresult}
    \expec{V} = T \, \frac{\sqrt{(1-h_{ren}^2)}}{\sqrt{2 \Lambda}}
\coth\left(\sqrt{2\Lambda(1-h_{ren}^2)}\,T\right).
\end{equation}
Dividing \myref{Nholeresultpartfunc} by \myref{averagevolresult} we find 
that the spacetime density $n$ of holes is constant,
\begin{equation}
\label{densityfinalresult}
n=\frac{\expec{N_\genus}}{\expec{V}}= \frac{h_{ren}^2}{1-h_{ren}^2} \,\Lambda.
\end{equation}
The density of holes in terms of the renormalized Newton's constant is given by
\begin{equation}
\label{densityonG}
n=\frac{1}{e^\frac{4\pi}{G_N}-1} \,\Lambda.
\end{equation}
The behaviour of $n$ in terms of the renormalized Newton's constant is shown 
in Fig.\ \ref{figure1}. The density of holes vanishes as $G_N\rightarrow 0$ and 
the model reduces to the case without topology change. -- An alternative calculation 
of the density of holes from an infinitesimal strip, which leads to the same result, 
is presented in Appendix B. 

\begin{figure}
\begin{center}
\includegraphics[width=4in]{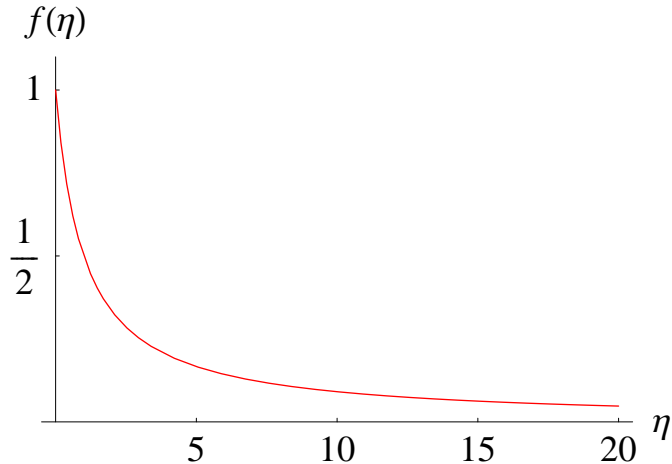}
\caption{The coefficient of the effective potential, $f(\eta)=1/(1+\eta)$,
as function of the density of holes in units of $\Lambda$, 
$\eta=\frac{n}{\Lambda}$.}\label{figure2}
\end{center}
\end{figure}

We can now rewrite and interpret the Hamiltonian \myref{Hamiltonian1} in 
terms of physical 
quantities, namely, the cosmological scale $\Lambda$ and the density of holes in 
units of $\Lambda$, i.e. $\eta=\frac{n}{\Lambda}$, resulting in
\begin{equation}
\label{hamiltonianeta}
\hat{H} (L,\dL)=-L\ddL-\dL+  \frac{1}{1+\eta} \,2\,\Lambda\,L.
\end{equation}
One sees explicitly that the topology fluctuations affect the dynamics since the 
effective potential depends on $\eta$, as illustrated by Fig.\ \ref{figure2}.

It should be clear from expressions \myref{hamiltonianeta} and \myref{densityonG} 
that the model has two scales instead of the single one of the pure
CDT model. As in the latter, 
the cosmological constant defines the global length scale of the
two-dimensional ``universe" through $\expec{L}\sim\frac{1}{\sqrt{\Lambda}}$. 
The new scale in the model with topology change is the relative scale $\eta$ 
between the cosmological and topological fluctuations, which is parametrized
by Newton's constant $G_N$. 

\section{Conclusions}\label{conclusions}

In this paper, we have presented the complete analytic solution
of a previously proposed model \cite{ourtopology1} of two-dimensional 
Lorentzian quantum gravity including a sum over topologies.
The presence of causality constraints imposed on the
path-integral histories -- physically motivated in
\cite{ourtopology1,ourtopology2} --
enabled us to derive a new class of continuum theories by taking an
unambiguously defined double-scaling limit of a statistical
model of simplicially regularized spacetimes. 
After computing the Laplace transform of the exact one-step
propagator of the discrete model, we investigated a 
two-parameter family, defined by (\ref{scalingh}) and (\ref{d}),  of
possible scalings for the gravitational (or Newton's) coupling,
from which physical considerations singled out a unique one.
For this case, we computed the quantum Hamiltonian, 
its spectrum and eigenfunctions, as well as the partition function and
propagator. Using these continuum results, we then calculated a
variety of physical observables, including the average spacetime 
density of holes and the expectation values of the spatial volume 
and all its moments.

This should be contrasted with the previous treatment in
\cite{ourtopology1,ourtopology2}, in which only the one-step
partition function with summed-over boundaries was evaluated. 
Because of the lack of boundary information, no explicit Hamiltonian
was obtained there. Moreover, it turns out that the scaling of the
couplings which in the current work led to the essentially unique
Hamiltonian (\ref{Hamiltonian1}) could {\it not} have been obtained
or even guessed
in the previous work. This is simply a consequence of the fact that
the dimensionful renormalized boundary cosmological constants
make an explicit appearance in the scaling relation (\ref{hm1}) for
$h$, and thus for Newton's constant. 
We conclude that -- unlike in the case of the original Lorentzian model -- for
two-dimensional causal quantum gravity with topology changes
one cannot obtain the correct scalings for the ``bulk" coupling
constants from the one-step partition function with boundaries
summed over (which is easier to compute than the full one-step
propagator). 

In contrast with what was extrapolated from the single-strip model in 
\cite{ourtopology2}, the total number of holes in a finite patch of 
spacetime turns out to be a finite 
quantity determined by the cosmological and Newton's constants.
Note that this finiteness result has been obtained dynamically and
without invoking any fundamental discreteness.
Since the density of holes is finite and every hole in the model is 
infinitesimal, this implies -- and is confirmed by explicit calculation -- that 
the expectation value of the number of holes in a general spatial slice 
of constant time is also
infinitesimal. The fact that physically sensible observables are
obtained in this toy model reiterates the earlier conclusion 
\cite{ourtopology1} that causality-inspired
methods can be a useful tool in constructing gravitational path
integrals which include a sum over topologies.

From the effective potential displayed in Fig.\ \ref{figure2} one observes 
that the presence of wormholes in our model leads to a decrease of the 
``effective" cosmological constant $f(\eta)\Lambda$. 
In Coleman's mechanism for
driving the cosmological constant $\Lambda$ to zero
\cite{Coleman:1988tj,Klebanov:1988eh}, an additional sum over different 
baby universes is performed in the path integral, which leads to a 
distribution of the cosmological constant that is peaked near zero. 
We do not consider such an additional sum over baby universes, 
but instead have an explicit expression for the effective potential which 
shows that an increase in the number of wormholes is accompanied by a 
decrease of the ``effective" cosmological constant. 
A first step in establishing whether an analogue of our suppression
mechanism also exists in higher dimensions would be to try and understand 
whether one can identify a class of causally preferred topology changes
which still leaves the sum over geometries exponentially bounded.
We will return to this issue in a future publication.

\section*{Acknowledgments}
S.Z. thanks J. Jers\'ak for valuable comments and acknowledges
support of the Dr. Carl Duisberg-Foundation
%: \textit{Auslandsstipendien f\"ur Studierende der Naturwissenschaften} 
and a Theoretical Physics Utrecht Scholarship.
R.L. acknowledges support by the Netherlands Organisation for
Scientific Research (NWO) under their VICI program.

\appendix
\section{Scalings with $\beta=1$}

In this appendix we discuss the scalings with $\beta=1$ which we
discarded as unphysical in Sec.\ \ref{continuumsection} above.
We proceed as before by inserting the scaling relations 
\myref{scalinga} and \myref{scalingbnew} into the composition law 
\myref{timeevolution1}. Instead of using $\beta=\frac{3}{2}$ we set $\beta=1$, 
leading to the scaling
\begin{equation}
%\label{ }
h=\frac{1}{4}\,h_{ren}\,a\,\sqrt{\Lambda}^\alpha (X+Y)^{1-\alpha},
\end{equation}
where the normalization factor on the right-hand side has been chosen 
for later convenience.
Up to first order in $a$ one obtains
\begin{equation}
(1-a \hat{H}+\mathcal{O}(a^2))\psi(X) =  \int^{i
\infty}_{-i\infty} \frac{dZ}{2\pi i} \left\{A(X,Z)+B(X,Z) a +
\mathcal{O}(a^2)\right\} \psi(Z),
\end{equation}
where the leading-order contribution is given by
\begin{equation}
%\label{ }
A(X,Z)=\frac{2}{(Z-X)\left(1+C(X,Z)\right)}
\end{equation}
with
\begin{equation}
%\label{ }
C(X,Z)=\sqrt{1-h_{ren}^2(X-Z)^{-2\alpha}\Lambda^\alpha}.
\end{equation}
For the Laplace transform of $A(X,Z)$ to yield a delta-function, the scaling 
should be chosen such that $\alpha\leqslant 0$. 
Considering now the terms of first order in $a$,
\begin{eqnarray}
%\label{ }
B(X,Z)&=&\frac{  h_{ren}^2(X+Z-4 Z \gamma ) 
\Lambda ^{\alpha }}{(X-Z)^{1+2\alpha} C(X,Z) \left(1+C(X,Z)\right)^2}\nonumber\\
&-&2\,\frac{X Z-2\Lambda + \gamma  (X-Z)^2 }
{(X-Z)^2 C(X,Z) \left(1+C(X,Z)\right)},\label{app_B}
\end{eqnarray}
one finds that for $\alpha\leqslant -1$ the continuum limit is independent 
of any ``hole contribution" (i.e. terms depending on $h_{ren}$) and therefore
leads to the usual Lorentzian model. This becomes clear when one expands 
the last term of \myref{app_B} in $(X-Z)$, resulting in
\begin{equation}
%\label{ }
\frac{X Z-2\Lambda }{(X-Z)^2 C \left(1+C\right)}
=\frac{1}{2}\frac{X Z-2\Lambda }{(X-Z)^2}\left(1+\frac{3}{4}\,h_{ren}^2 \Lambda^\alpha
(X-Z)^{-2\alpha}+\mathcal{O}((X-Z)^{-4\alpha})\right).
\end{equation}
For $\alpha\leqslant -1$ the term depending on $h_{ren}$ does not have a pole 
and therefore does not contribute to the Hamiltonian.
Since we are only interested in non-fractional poles, this leaves as possible
$\alpha$-values only $\alpha=0$ and $\alpha=-\frac{1}{2}$.

\subsection{The case $\beta =1$, $\alpha=0$}

For $\alpha=0$ the Hamiltonian retains a $\gamma$-dependence contained
in the first line of \myref{app_B}. 
Since there is no immediate physical
interpretation of $\gamma$ in our model, it seems natural to choose
$\gamma=0$, although strictly speaking this does not resolve the
problem of explaining the $\gamma$-dependence of the continuum limit.
Setting this question aside, one may simply look at the resulting model
as an interesting integrable model in its own right. 
In order to obtain a delta-function to leading order, one still needs
to normalize the transfer matrix by a constant factor $2/(1+s)$,
with $s:=\sqrt{1-h_{ren}^2}$.
After setting $\gamma=0$ and performing an inverse Laplace transformation, the
Hamiltonian reads
\begin{equation}
\label{haml}
\hat{H}(L,\dL)=\frac{1}{s}\left(-L\ddL -s\dL +2\Lambda L \right).
\end{equation}
It is self-adjoint with respect to the measure $d\mu(L)=L^{s-1}dL$. 
Further setting $L=\frac{\varphi^2}{2\,s}$ one encounters the one-dimensional
Calogero Hamiltonian
\begin{equation}
\label{calo}
\hat{H}(\varphi,\frac{\partial}{\partial\varphi})=-\frac{1}{2}\frac{\partial^2}{\partial\varphi^2}+
\frac{1}{2}\omega^2\varphi^2-\frac{1}{8}\frac{A}{\varphi^2},
\end{equation}
with $\omega\equ\frac{\sqrt{2 \Lambda}}{s}$ and $A\equ 1-4(1-s)^2$, which implies 
that the model covers the parameter range 
$-3\leqslant A\leqslant 1$. The maximal range for which the Calogero Hamiltonian is 
self-adjoint is $-\infty<A\leqslant 1$. The usual Lorentzian model without holes corresponds 
to $A\equ 1$. The Hamiltonian (\ref{calo}) has already appeared in a causal dynamically
triangulated model where the two-dimensional geometries were decorated with a
certain type of ``outgrowth" or small ``baby universes" \cite{DiFrancesco:2000nn}.
This model covered the parameter range $0\leqslant A\leqslant 1$.

The eigenvectors of the Hamiltonian (\ref{haml}) are given by
\begin{equation}
%\label{ }
\psi_n(L)=\mathcal{A}_n e^{- \sqrt{2\Lambda}L}\,{}_1F_1(-n,s,2\sqrt{2\Lambda} L),
\quad d\mu(L)=L^{s-1}dL,
\end{equation}
where ${}_1F_1(-n,a,b)$ is the Kummer confluent hypergeometric function. 
The eigenvectors form an orthonormal basis with the normalization factors
\begin{equation}
%\label{ }
\mathcal{A}_n=(8\Lambda)^{\frac{s}{4}}\sqrt{\frac{\Gamma(n+s)}{\Gamma(n+1)
\Gamma(s)^2}}
\end{equation}
and the corresponding eigenvalues
\begin{equation}
%\label{ }
E_n=\frac{\sqrt{2\Lambda}}{s}(2n+s),\quad n=0,1,2,...\, .
\end{equation}
One sees explicitly that the case $s\equ 1$ or, equivalently,
$A\equ 1$ corresponds to the pure two-dimensional CDT model.

\subsection{The case $\beta =1$, $\alpha=-\frac{1}{2}$}

For $\alpha=-\frac{1}{2}$ the result does not depend on $\gamma$ and therefore
on the detailed manner in which we approach the critical point. However, 
the Hamiltonian
\begin{equation}
%\label{ }
\hat{H}(L,\dL)=-L\ddL-\dL+2\,\Lambda\,L-\frac{3}{4}\,h_{ren}^2\Lambda^{-1/2}\,\ddL
\end{equation}
cannot be made self-adjoint with respect to any measure $d\mu(L)$ because the 
boundary part of the partial integration always gives a nonvanishing contribution. 
We therefore discard this possibility.

\section{The density of holes of an infinitesimal strip}

In this appendix we give an alternative derivation of the spacetime density $n$ of holes 
and explicitly show that the number of holes in a spacetime strip of infinitesimal time 
duration $a$ is also infinitesimal.
The operator in the $L$-representation of the number of holes per infinitesimal strip 
with fixed initial boundary $L$ can be calculated by
\begin{equation}
\label{N1}
\hat{N}_{\genus,a\rightarrow 0} =\hat{T}^{-1}\frac{h_{ren}}{2}\frac{\partial\, \hat{T}}{\partial h_{ren}},
\end{equation}
where $\hat{T}$ is the transfer matrix defined in \myref{transfer}. 
Using $\hat{T}=1-a\hat{H}+\mathcal{O}(a^2)$ and evaluating \myref{N1} to leading order
in $a$ gives
\begin{equation}
\label{N2}
\hat{N}_{\genus,a\rightarrow 0}=-a\,\frac{h_{ren}}{2}\frac{\partial \hat{H}}{\partial h_{ren}}+
\mathcal{O}(a^2)=2\, \Lambda\,h_{ren}^2\,L\,a+\mathcal{O}(a^2).
\end{equation}
Similarly, the volume operator of the same infinitesimal spacetime strip in the $L$-representation
is given by 
\begin{equation}
\label{volop}
\hat{V}_{a\rightarrow 0} =-\hat{T}^{-1} \frac{\partial\, \hat{T}}{\partial \Lambda}=
a\,\frac{\partial \hat{H}}{\partial \Lambda}+\mathcal{O}(a^2)=
2\, (1-h_{ren}^2)\,L\,a+\mathcal{O}(a^2).
\end{equation}
Although both expressions (\ref{N2}) and (\ref{volop}) vanish in the limit as $a\rightarrow 0$
(and therefore the number of holes and the strip volume are both ``infinitesimal"),
their quotient evaluates to a finite number independent of $L$, namely,
\begin{equation}
n=\frac{N_{\genus,a\rightarrow 0}}{V_{a\rightarrow 0}}= \frac{h_{ren}^2}{1-h_{ren}^2} \,\Lambda.
\end{equation}
This is the exactly the same result for the spacetime density $n$ of holes as we obtained 
earlier from the continuum partition function \myref{densityfinalresult}.

\vspace{1cm}
%\pagebreak

\end{document}